# Determination of the quark-gluon string parameters from the data on pp, pA and AA collisions at wide energy range using Bayesian Gaussian Process Optimization

**Vladimir Kovalenko[1]**
*Saint Petersburg State University*
*Universitetskaya nab., 7-9, 199034, St. Petersburg, Russia*
*E-mail:* `v.kovalenko@spbu.ru`

Bayesian Gaussian Process Optimization can be considered as a method of the determination of the model parameters, based on the experimental data. In the range of soft QCD physics, the processes of hadron and nuclear interactions require using phenomenological models containing many parameters. In order to minimize the computation time, the model predictions can be parameterized using Gaussian Process regression, and then provide the input to the Bayesian Optimization. In this paper, the Bayesian Gaussian Process Optimization has been applied to the Monte Carlo model with string fusion. The parameters of the model are determined using experimental data on multiplicity and cross section of pp, pA and AA collisions at wide energy range. The results provide important constraints on the transverse radius of the quark-gluon string ($r_{str}$) and the mean multiplicity per rapidity from one string ($\mu_0$).

---

[1]Speaker

# 1. Introduction

One of the new modern methods that can be used to optimize model parameters on the basis of experimental data is the Bayesian approach [1, 2, 3]. This approach allows us to determine the regions of admissible (on the basis of available experimental data) model parameters with allowance for possible correlations between the parameters.

In view of the complexity of the evolution of nucleus-nucleus collisions and the need for large statistics, taking into account the scanning by more parameters, modern Monte Carlo calculations require large computational resources [4, 5]. To interpolate model calculations for intermediate parameter values, the so-called Gaussian Processes for Machine Learning [1] can be used, allowing to reduce the time required for computation by orders of magnitude.

In this paper, the Gaussian process was applied to the Monte Carlo model with string fusion [6, 7, 8] to optimize a number of parameters based on multiplicity data in pp, p-Pb and Pb-Pb collisions at LHC energies. This model enables to describe collisions with protons and nuclei on the partonic level. The main feature of the model is that it allows color tubes (strings) to fuse with each other and form strings with higher tension.

The following parameters were considered for the optimization: the radius of the nucleon, the confinement radius, the effective dipole coupling constant, the mean multiplicity per rapidity unit from one string ($\mu_0$), the transverse radius of the string ($r_{str}$).

The calculations were made as part of the GPML package Matlab Code, v. 4.1 [9], in the GNU Octave computing environment, v. 4.0.0 [10]. We used a quadratic exponent as a correlation function and a Gaussian likelihood function. The results were cross-checked in Scikit-learn Gaussian Process Regressor (python 2.7.12, Sklearn 0.18.1) [11].

# 2. Bayesian Gaussian process for model parameters optimization

The typical schema of the framework of the parameter tuning using Bayesian Gaussian process is shown in the figure 1.

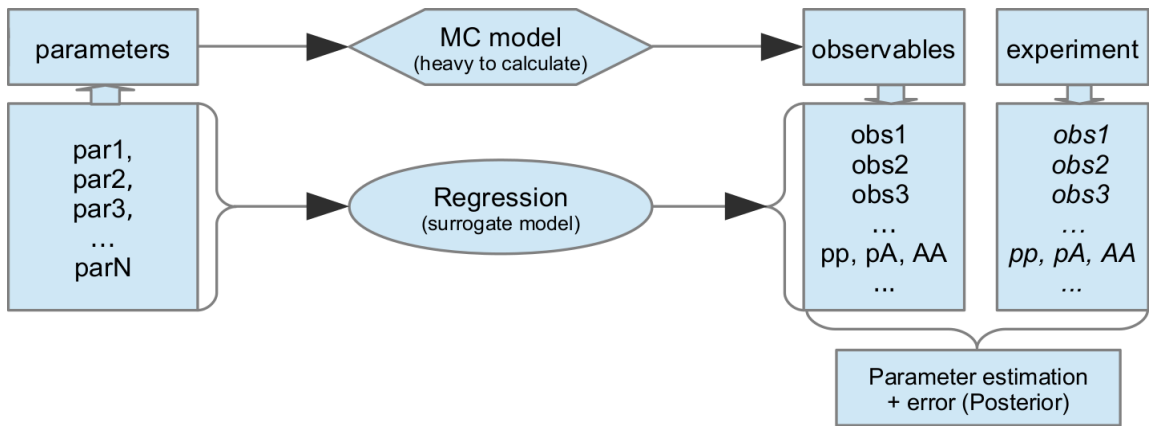

**Figure 1:** Application of Bayesian Gaussian process for model parameters optimization.

The heavy demanding model (labeled here as MC model) takes as input the set of the parameters and produces in the output the definite set of observables. For example, in heavy ion collisions the observables can be: transverse momentum spectra of the charged particles, pt-dependent flow coefficients, particle yields ratios etc.

Instead of the full model, the heavy calculations are replaced by a surrogate model (trained regression). For this purpose, the Professor software [12-14] can be used, which provides multi-variable polynomial interpolation. However, it is not flexible enough and still quite demanding. The Gaussian Process parametrization appears as a good alternative. In this approach effectively the fitting function appears as a Gauss function with specific kernel.

## 2.1 Stochastic Gaussian process.

Stochastic process is Gaussian if and only if for every finite set of indices $\mathbf{X}_{t_1,\ldots,t_k} = (\mathbf{X}_{t_1},\ldots,\mathbf{X}_{t_k})$ is a multivariate Gaussian random variable.
Gaussian processes can be completely defined by covariance functions. For example:

- Constant : $K_{\mathrm{C}}(x, x') = C$
- Linear: $K_{\mathrm{L}}(x, x') = x^T x'$
- Gaussian noise: $K_{\mathrm{GN}}(x, x') = \sigma^2 \delta_{x,x'}$
- Squared exponential: $K_{\mathrm{SE}}(x, x') = \exp\left(-\frac{\|d\|^2}{2\ell^2}\right)$

In our analysis we will use sum of Squared exponential and Gaussian noise:

$$k(\mathbf{x}_i, \mathbf{x}_j) = \exp\left(-\frac{|\mathbf{x}_i - \mathbf{x}_j|^2}{2\ell^2}\right) + \sigma_n^2 \delta_{ij} \qquad (1)$$

The general advantages of Gaussian processes are the following [11]:
– The prediction interpolates the observations (at least for regular kernels).
– The prediction is probabilistic (Gaussian), one can compute empirical confidence
intervals - can be used for online fitting, adaptive refitting

Disadvantages of Gaussian processes:
– They are not sparse, i.e., they use the whole samples/features information to
perform the prediction.
– They lose efficiency in high dimensional spaces – namely when the number of
features exceeds a few dozens.

In our analysis thees disadvantages are not crucial, because we had quite a low number of observables (features).

## 2.2 Bayesian parameters optimization.

After the model is defined, its prediction can be extracted with uncertainty and the comparison with the experimental data is characterized by likelihood.

According to Bayes' theorem:

$$\text{posterior} = \frac{\text{likelihood} \times \text{prior}}{\text{marginal likelihood}}, \qquad p(x|\mathbf{y}, X) = \frac{p(\mathbf{y}|X, x)p(x)}{p(\mathbf{y}|X)} \qquad (2)$$

For the prior distribution we take uniform one. If the model is good in constraining the data, the posterior will very slightly depend on prior.

$$P(\mathbf{x}) \propto \begin{cases} 1 & \text{if } \min(x_i) \leq x_i \leq \max(x_i) \text{ for all } i \\ 0 & \text{else.} \end{cases} \quad (3)$$

The marginal likelihood in formula (2) appears as a normalization constant. Hence, the likelihood (determined from the Gaussian Processes)

$$\ln P(x|y,X) = -\frac{1}{2}\frac{(y-y_{\exp})^2}{\sigma_{\exp}^2} + const, \quad (4)$$

and the Posterior

$$P(x|y,X) \propto e^{-\frac{1}{2}\frac{(y-y_{\exp})^2}{\sigma_{\exp}^2}}. \quad (5)$$

For several observables, we take independent Gaussian Process for each one and then combine in the Likelihood (as a sum). The more advanced technique would be applying here the Principal Component Analysis to observables in order to exclude inter-correlations among different observables.

## 3. Monte Carlo model description

The present model [6, 7, 8] is based on the partonic picture of nucleon collisions. Each nucleon is supposed to consist of a valence quark-diquark pair and a certain number of sea quark-antiquark pairs (see Fig. 1). The number of pairs is distributed according to Poisson low. The total momentum of a nucleon is shared between partons according to exclusive distribution [6]:

$$\rho(x_1,\ldots x_N) = c \cdot \prod_{j=1}^{N-1} x_j^{\frac{-1}{2}} \cdot x_N^{\alpha_N} \cdot \delta\left(\sum_{i=1}^{N} x_i - 1\right), \quad (6)$$

Here the valence quark is labeled by $N-1$, diquark – $N$, and the rest numbers correspond to quark-diquark pairs; $\alpha_N = 3/2$ with probability of $2/3$ (ud-diquark configuration), and $\alpha_N = 5/2$ with probability of $1/3$ (uu-diquark configuration).

An elementary interaction is realized in the model of color dipoles. The transverse coordinates of the dipoles are generated according to Gaussian distribution, with mean-squared transverse radius $r_0 = \sqrt{\frac{2}{3}} r_N$, where $r_N$ is a nucleon radius. It is assumed that each quark-diquark and quark-antiquark pair forms a dipole; and the probability amplitude of the collision of two dipoles from target and projectile is given by:

$$f = \frac{\alpha_s^2}{2} \ln^2 \frac{|\vec{r}_1 - \vec{r}_1'||\vec{r}_2 - \vec{r}_2'|}{|\vec{r}_1 - \vec{r}_2'||\vec{r}_2 - \vec{r}_1'|}, \quad (7)$$

where $(\vec{r}_1, \vec{r}_2), (\vec{r}_1', \vec{r}_2')$ are transverse coordinates of the projectile and target dipoles, and $\alpha_s$ – come effective constant, which value is a tunable parameter of the model. After taking into account the confinement effects the probability amplitude is:

$$f = \frac{\alpha_s^2}{2}[K_0(\frac{|\vec{r}_1-\vec{r}_1'|}{r_{max}}) + K_0(\frac{|\vec{r}_2-\vec{r}_2'|}{r_{max}}) - K_0(\frac{|\vec{r}_1-\vec{r}_2'|}{r_{max}}) - K_0(\frac{|\vec{r}_2-\vec{r}_1'|}{r_{max}})]^2. \qquad (8)$$

Here $r_{max}$ is characteristic confinement scale.

Note that according to these formulas (7) – (8), two dipoles interact more probably, if the ends are close to each other, and (others equal) if they are wide. If in Monte Carlo simulation there is a collision between two dipoles, two quark-gluon strings are stretched between the ends of the dipoles, and the process of string fragmentation gives observable particles.

An important feature of the model is that each dipole can participate in inelastic collision with color string formation only once, in order to keep the energy conservation in an elementary nucleons collision [15, 16].

Multiplicity and transverse momentum are calculated in the approach of color strings, stretched between projectile and target partons, taking into account their finite rapidity width. In the calculation of the multiplicity, the interaction between several strings in the transverse plane is taken into account, which is performed in the model of string fusion [17, 18, 19], according to which multiplicity and transverse momentum from a cluster of overlapped strings:

$$\langle \mu \rangle_k = \mu_0 \sqrt{k} \frac{S_k}{\sigma_0} \quad \langle p_t^2 \rangle = p_0 \sqrt{k}, \qquad (9)$$

where $S_k$ – area, where $k$ strings are overlapping, $\sigma_0 = \pi r_{str}^2$ – single string transverse area, $\mu_0$ and $p_0$ – mean multiplicity and transverse momentum from one string.

Note that the amount of string fusion effect depends on the transverse radius of string: the thicker the strings, the more of them are overlapping. The limit of non-interacting strings corresponds to zero string radius.

## 4. Parameter dependence approximation

The main parameters of the model are the following:
– mean number of dipoles $\lambda$ ;
– transverse mean-squared radius of nucleon $r_0$ ;
– confinement scale $r_{max}$ ;
– transverse radius of string $r_{str}$ ;
– mean multiplicity from single string per unit of rapidity $\mu_0$ .

It is assumed that only $\lambda$ depends on collision energy $\sqrt{s}$ .

For each parameter set, value of $\lambda$ is obtained using the correspondence between mean number of dipoles and collision energy $\lambda = \lambda(\sqrt{s}; r_0, r_{max})$ . This step does not include the calculation of multiplicity and, hence, does not involve $r_{str}$ and $\mu_0$ values. For the rest, the Bayesian Gaussian Process is applied.

We start from uniform priors in the range:
r $_0$ : 0.4 – 0.7 fm;

$r_{max}/r_0 : 0.3 – 0.6$;
$\alpha_s : 0.2 – 2.8$;
$r_{str} : 0 – 0.6$ fm;

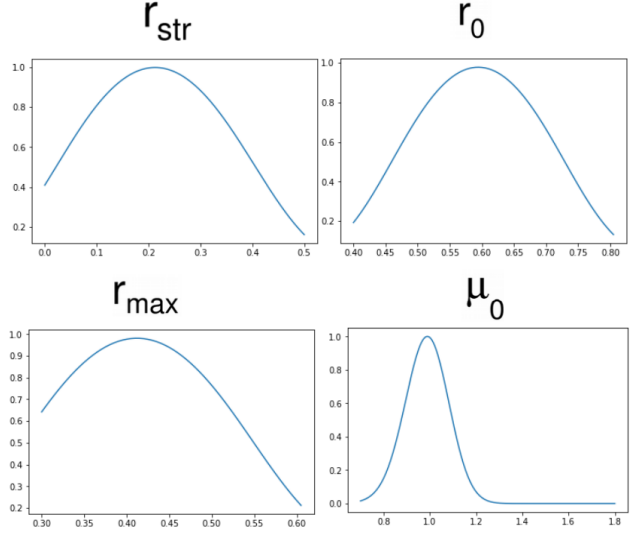

**Figure 2:** Posterior distributions after accounting of pp multiplicity.

Figure 2 shows the posterior distribution after according of pp multiplicity in a wide energy range (900 GeV – 7 TeV). We see that this observable (pp multiplicity) is not much sensitive to string fusion $r_{str}$, however, it favors $r_0$ around 0.6 fm. Using only this data, $r_{max}$ and $\alpha_s$ appear not well restricted. $\mu_0$ using this data clearly peaks around 1.0.

Figure 3 shows the likelihood for posterior parameter estimation from energy dependence of pp multiplicity. It shows that both energies restrict the parameters in a consistent way.

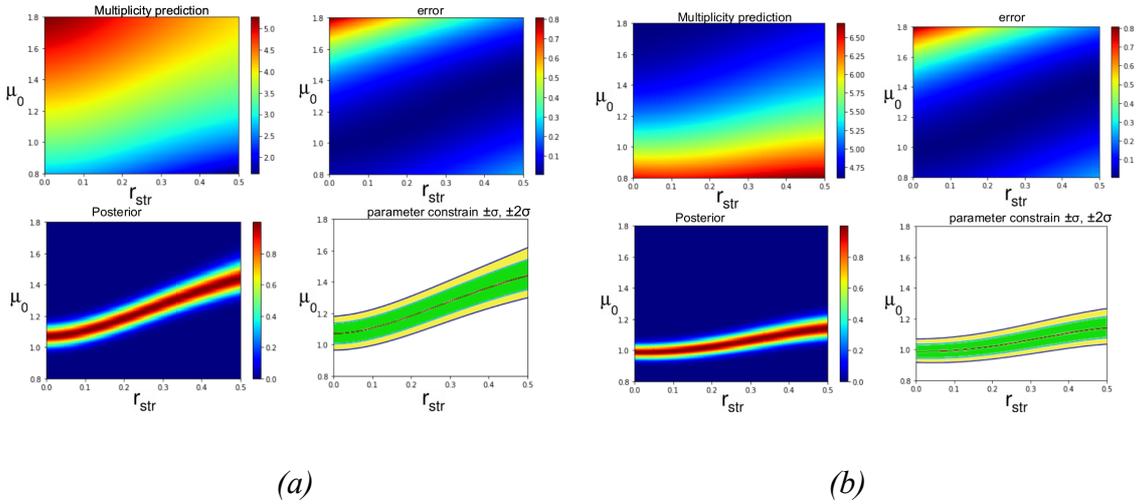

*(a)*                                              *(b)*

**Figure 3:** Posterior parameter estimation from energy dependence of pp multiplicity (*a*: using pp interaction at $\sqrt{s}=0.9$ TeV, *b*: at $\sqrt{s}=7$ TeV).

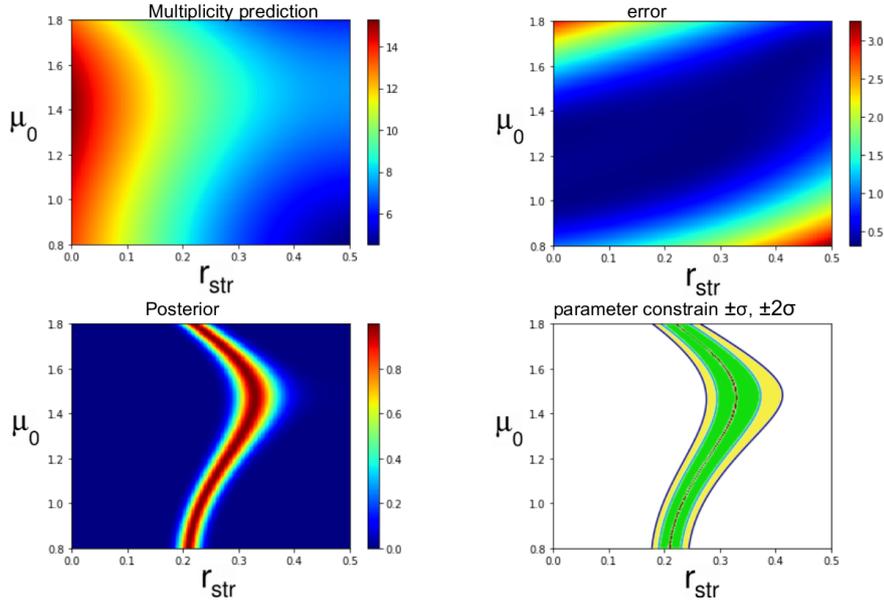

**Figure 4:** Posterior parameter estimation from central PbPb collisions at 2.76 TeV.

In figure 4 the posterior parameter estimation is shown after taking into account the data on multiplicity from central PbPb collisions at 2.76 TeV. We see that heavy ion data helps a lot to restrict the transverse string radius $r_{str}$. For the most values of other parameters, $r_{str}$ is restricted within 0.2-0.3 fm.

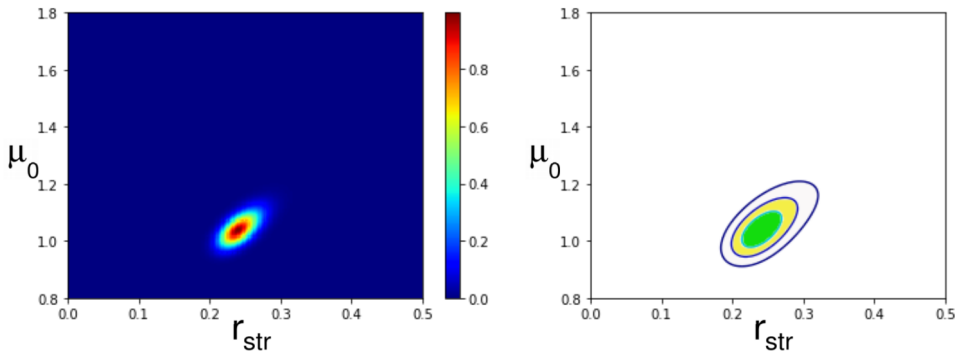

**Figure 5:** Combined results on Posterior from the data on pp multiplicity in a wide energy range together with data on multiplicity in central Pb-Pb collisions at 2.76 TeV. Contour plot in the right panel shows 1, 2, 3 sigma levels.

In figure 5 the combined posterior distribution is shown for all the data used. We obtained the following final estimation:

$$r_{str} = 0.25 \pm 0.03 \, \text{fm}$$
$$\mu_0 = 1.1 \pm 0.03$$

## 5. Conclusions

Bayesian Gaussian process optimization has been applied for the parameter tuning of the non-Glauber Monte Carlo model with string fusion. In the model, the inelastic cross section and multiplicity are described over wide energy range and for different colliding systems. Multiplicity per rapidity from one single string $\mu_0$ is constrained by the energy dependence of the multiplicity in pp collisions. The transverse radius of string $r_{str}$ (string fusion parameter) is constrained by multiplicity in central PbPb collisions. Using these data we obtained the final estimation: $r_{str}=0.25\pm0.03$ fm, $\mu_0=1.1\pm0.03$.

The possible improvement of the framework and the parameter estimation procedure can be done by considering more data, extension of the energy range and application of the Principal Component Analysis to all the observables.

## 6. Acknowledgments

The research was supported by Russian Science Foundation under grant 17-72-20045.

## References

[1] C. E. Rasmussen, C. K. I. Williams, *Gaussian Processes for Machine Learning*. The MIT Press, 2006

[2] Jonah E. Bernhard, et al, Phys. Rev. C 94, 024907 (2016)

[3] Jonah E. Bernhard, arXiv:1804.06469 [nucl-th], Ph.D. dissertation (2018)

[4] I. Altsybeev, et al, GRID-2012, Dubna, JINR, pp. 18-22, 2012

[5] A. Bogdanov, et al, GRID-2012, Dubna, JINR, pp.164-168, 2012

[6] V. N. Kovalenko. Phys. Atom. Nucl. 76, 1189 (2013), arXiv:1211.6209 [hep-ph]

[7] V. Kovalenko, V. Vechernin, PoS (Baldin ISHEPP XXI) 077, arXiv:1212.2590 [nucl-th], 2012

[8] V. Kovalenko, Kovalenko, PoS QFTHEP2013 (2013) 052

[9] GPML Matlab Code http://www.gaussianprocess.org/gpml/code/matlab/doc/

[10] GNU Octave https://www.gnu.org/software/octave

[11] Scikit-Learn Gaussian Process Regressor http://scikit-learn.org/stable/modules/generated/sklearn.gaussian_process.GaussianProcessRegressor.html

[12] Professor https://professor.hepforge.org/

[13] P. Abreu, et al (DELPHI Collaboration), Z. Phys. C 73, 11-6 (1996)

[14] A. Buckley, et al, Eur. Phys. J. C 65, 331-357 (2010), arXiv:0907.2973 [hep-ph]

[15] V. N. Kovalenko, arXiv:1308.1932 [hep-ph] (2013)

[16]  T. Drozhzhova, G. Feofilov, V. Kovalenko, A. Seryakov, PoS (QFTHEP 2013) 052, 2013

[17]  V. V. Vechernin, R. S. Kolevatov, Phys. Atom. Nucl. 70, 1797 (2007)

[18]  M. A. Braun, C. Pajares, V. V. Vechernin, Phys. Let. B 493, 54 (2000)

[19]  M. Braun, R. Kolevatov, C. Pajares, V. Vechernin, Eur. Phys. J. C 32, 535 (2004)